\documentclass{optica-article}

\journal{opticajournal} % for journals or Optica Open

\articletype{Research Article}

\usepackage{lineno}
\usepackage[version=4]{mhchem}
\begin{document}

\title{Designing metallo-dielectric antennas for cryogenic applications}

%\affil[1]{\orgname{Max Planck Institute for the Science of Light}, \orgaddress{\postcode{91058} \city{Erlangen}, \country{Germany}}}

%\affil[2]{\orgdiv{Department of Physics}, \orgname{Friedrich Alexander University Erlangen-Nuremberg}, \orgaddress{\postcode{91058} \city{Erlangen}, \country{Germany}}}

%\affil[3]{\orgdiv{Graduate School in Advanced Optical Technologies (SAOT)}, \orgname{Friedrich Alexander University Erlangen-Nuremberg}, \orgaddress{\postcode{91058} \city{Erlangen}, \country{Germany}}}

\author{Siwei Luo,\authormark{1,2} Tim Hebenstreit,\authormark{1,2} Alexey Shkarin, \authormark{1} Jan Renger,\authormark{1} Tobias Utikal, \authormark{1} and Stephan Götzinger \authormark{1,2,3,4}}

\address{\authormark{1} Max Planck Institute for the Science of Light, 91058 Erlangen, Germany\\
\authormark{2} Department of Physics, Friedrich Alexander University Erlangen-Nuremberg, 91058 Erlangen, Germany\\
\authormark{3} Graduate School in Advanced Optical Technologies (SAOT), Friedrich Alexander University Erlangen-Nuremberg, 91058 Erlangen, Germany\\
\authormark{4} stephan.goetzinger@mpl.mpg.de}

%\email{\authormark{*}opex@optica.org} %% email address is required; see note below about the corresponding author designation

% use {asbstract*} to suppress the copyright line. Copyright information will be added in production

\begin{abstract*} 

We present the design of cryogenic metallo-dielectric antennas tailored to single organic emitters, where the choice of host material imposes specific constraints on the antenna geometry. Using dibenzoterrylene (DBT) in para-dichlorobenzene (\ce{\textit{p}-DCB}) as a model system, we show that photon collection efficiencies exceeding $90\,\%$ can be achieved for arbitrary dipole orientations of the fluorescent molecule. The antenna design is intrinsically broadband and tolerant to emitter positioning within the structure. We further provide concrete fabrication guidelines for the full antenna architecture, and experimentally demonstrate its operating principle by recording a back-focal-plane image (BFP) of a single molecule inside a fabricated antenna.

% This template contains important information on submissions to Optica Publishing Group journals. We encourage authors to focus foremost on their content and not on formatting. The template is provided as a guide, but following the visual styling is optional. Each manuscript will be formatted in a consistent way during production. Authors also have the option to \href{https://opticaopen.org}{submit articles} to the Optica Publishing Group preprint server, \href{https://opticaopen.org/figshare}{Optica Open}. You may find it helpful to use our optional \href{https://preflight.paperpal.com/partner/optica/$opticapublishinggroupjournals}{Paperpal manuscript readiness check}  and \href{https://$languageediting.optica.org/}{language polishing service}. Note that copyright and licensing information should not be added to your journal or Optica Open manuscript.

\end{abstract*}

\section{Introduction}
Single-photon sources are essential building blocks for quantum technology applications, including quantum key distribution \cite{gisin2002quantum}, linear optical quantum computing \cite{knill2001scheme,kok2007linear}, and quantum networks \cite{kimble2008quantum,sangouard2011quantum}. Recent years have seen rapid progress toward sources combining high purity, near-unity photon indistinguishability, and efficient photon extraction \cite{somaschi2016near,ding2025high}. Much of this progress relies on embedding a single emitter in an optical microcavity \cite{ding2025high} or coupling it to a waveguide \cite{uppu2020scalable}, thereby enhancing collection efficiency through Purcell-enhanced emission into a well-defined mode. Despite this progress, there is still room for improvements concerning the collection efficiency, limiting the practical deployment of these sources.

An alternative route to efficient photon collection is provided by planar metallo-dielectric antennas, which at room temperature have demonstrated photon collection efficiencies exceeding $95\,\%$ \cite{lee2011planar,chu2014experimental}. Such antennas are appealing because they are broadband and largely insensitive to the emitter's exact position and dipole orientation. The insensitivity to position and orientation is especially valuable for emitters whose location and dipole axis within a host material cannot be deterministically controlled, in contrast to platforms such as lithographically positioned quantum dots \cite{dousse2008controlled,gschrey2013situ}. However, photon indistinguishability, a prerequisite for most quantum photonic protocols, requires cryogenic operation. Realizing such a design at cryogenic temperature is nontrivial: the room-temperature demonstrations rely on large-numerical-aperture (NA) oil-immersion objectives, which cannot be used inside a cryostat, so one is restricted to air objectives with substantially lower NA. Compensating for this reduced collection angle requires re-optimizing the antenna's refractive-index contrast and layer materials, so that its performance is preserved while remaining compatible with cryogenic operation.

Single organic molecules embedded in molecular crystals are excellent solid-state quantum emitters, offering lifetime-limited optical transitions, high photostability, and near-unity quantum efficiency at cryogenic temperatures. Systems such as DBT in aromatic host crystals are particularly attractive as bright, coherent single-photon sources for quantum optics and quantum information processing \cite{nicolet2007single,verhart2016spectroscopy}. Realizing their full potential, however, hinges on efficiently collecting the emitted photons. Previous approaches to collecting photons from organic molecules have reported impressive collection efficiencies of up to $40\,\%$ \cite{colautti2020a3d}, but still fall short of the near-unity efficiencies required for many quantum photonic applications. A photon collection methodology that is tailored to organic molecules, compatible with cryogenic operation, and able to approach near-unity collection efficiency is so far missing.

The paper is structured as follows: we present the design principles for a cryogenic metallo-dielectric antenna tailored to single organic emitters, using DBT in \ce{\textit{p}-DCB} as a model system. Combining an analytical multilayer model \cite{chen2007efficient,chen201199} with finite-difference time-domain (FDTD) simulations, we show that photon collection efficiencies exceeding $90\,\%$ can be reached for arbitrary dipole orientations using a collection lens of \text{NA}=0.77, provided the organic layer is kept sufficiently thin. Building on these principles, we present three specific antenna geometries and analyze their performance in detail. We then translate this framework into a complete fabrication workflow -- comprising two-step lithography, direct bonding of a solid immersion lens, and controlled melt growth of a 140-nm-thick molecular crystal -- and confirm via BFP imaging the operation of the antenna.
%The paper is structured as follows: we first introduce the analytical multilayer model and FDTD simulations used to derive the antenna's guiding principles, then present and compare three specific implementations, describe the fabrication workflow in detail, and finally present a back-focal-plane image of a single molecule inside an at cryogenic temperatures.

%%%%%%%%%%%%%%%%%%%%%%%%%%  body  %%%%%%%%%%%%%%%%%%%%%%%%%%

\section{Antenna design} \label{sec_simulation}

In this section, we analyze the dipole radiation pattern of different antenna schemes using the analytical model of Refs.~\cite{chen2007efficient, chen201199} and FDTD simulations performed with Ansys Lumerical FDTD. The analytical model computes the radiated power of a dipole embedded in an arbitrary planar multilayer stack, while the FDTD simulations additionally capture the full near- and far-field distribution. In both cases, horizontal (HED) and vertical (VED) electric dipole orientations are evaluated independently, and the photon collection efficiency is obtained by normalizing the angular power density distribution and integrating it over the collection solid angle defined by the objective's NA. The results underpin a discussion of the physical mechanisms governing the antenna design and its parameter optimization.

The concept of the metallo-dielectric antenna was introduced several years ago~\cite{lee2011planar}; we follow these design principles and adapt them for cryogenic applications. The simplest antenna design comprises a three-layer stack with a decreasing refractive-index profile $n_1>n_2>n_3$, where $n_1$ is a high-index substrate, $n_2$ is the layer containing the emitter, and $n_3$ is a low-index bottom layer. This configuration directs emission preferentially toward the high-index substrate, where the interface supports leakage into the substrate rather than being totally internally reflected. To minimize losses into the low-index half-space, a metallic reflector is placed below the low-index layer to redirect photons that would otherwise propagate in undesired directions, completing a four-layer antenna structure.

At room temperature, large-NA oil-immersion objectives relax the required refractive-index contrast between the emitter layer and the high-index substrate. In a cryostat, however, one is typically restricted to simpler air objectives with NA up to about $0.8$; for what follows, we specifically choose an aspheric lens with an NA of $0.77$. Because of the air gap between the antenna and the collection lens, we use a cubic zirconia solid immersion lens (SIL) as the high-index substrate layer. The zirconia SIL offers two advantages: first, its high refractive index of $n = 2.14$ increases the accessible wavevector range; second, its spherical surface allows photons to cross the interface at normal incidence, avoiding the additional refraction that would otherwise increase the NA required for efficient photon collection. Note that the curved surface of the SIL needs an anti-reflection coating, since the uncoated Fresnel reflectance at normal incidence of $\approx 13\,\%$ is otherwise non-negligible.

The organic layer containing the fluorescent molecules acts as a thin dielectric quasi-waveguide. Its birefringence introduces an additional design constraint: typical organic host crystals have a refractive index exceeding $1.8$ along at least one principal axis, and our host material, \ce{\textit{p}-DCB}, is no exception, with a reported value of $1.9$ \cite{manghi1967optical}. This value, however, was measured at $\lambda=589\,\mathrm{nm}$ \cite{manghi1967optical}, and the dispersion of the refractive index toward the DBT's emission wavelength has, to our knowledge, not been characterized. Because the crystal axes of \ce{\textit{p}-DCB} during growth are not controlled, the largest of these principal refractive indices must be assumed. Once the organic layer thickness becomes comparable to the emission wavelength of about $744\,\mathrm{nm}$, it begins to behave as a slab waveguide whose effective mode index of the lowest-order mode approaches this large bulk value, thereby reducing the contrast of refractive indices with the zirconia substrate. Since the wavevector component parallel to the interface is conserved across each layer, a reduced index contrast leads to emission angles too large to be captured by the collection lens, so that a high collection efficiency can no longer be guaranteed.

\begin{figure}[!htbp]
   \centering
   \includegraphics{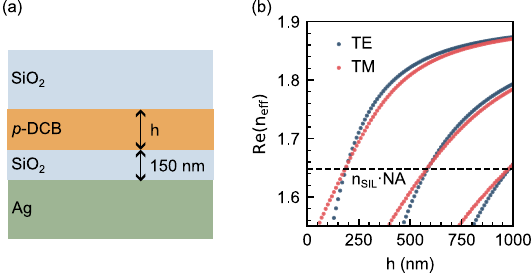}
 \caption{\textbf{Reducing the effective mode index of the organic layer.} (a) Schematic of the structure used in the simulation, representing the relevant waveguiding components in the antenna; see main text for details. (b) Real part of the effective mode index as a function of \textit{p}-DCB layer thickness. For thicknesses below approximately $200\,\text{nm}$, the mode index drops below $1.65$, the threshold required for efficient photon collection with the SIL and a collection lens of NA $= 0.77$.}
    \label{fig_antenna_introduction_4}
\end{figure}

This effect can be mitigated by keeping the organic quasi-waveguide thin enough to lower its effective index. We therefore performed FDTD simulations to determine the effective index of a \textit{p}-DCB planar waveguide as a function of thickness, assuming a bulk refractive index of 1.9. Fig.~\ref{fig_antenna_introduction_4}(a) shows the structure used for this simulation; we omit the cubic zirconia high-index layer here for simplicity, since it affects mostly the imaginary part of the effective mode index. As our focus is on reducing the real part, this simplification does not compromise the validity of our results. As shown in Fig.~\ref{fig_antenna_introduction_4}(b), the effective indices of both lowest-order modes fall below the required threshold of 1.65 for thicknesses below approximately $200\,\mathrm{nm}$, ensuring that our specific SIL--aspheric-lens combination can capture the extracted photons for HED and VED.

\begin{figure}[!htbp]
   \centering
   \includegraphics{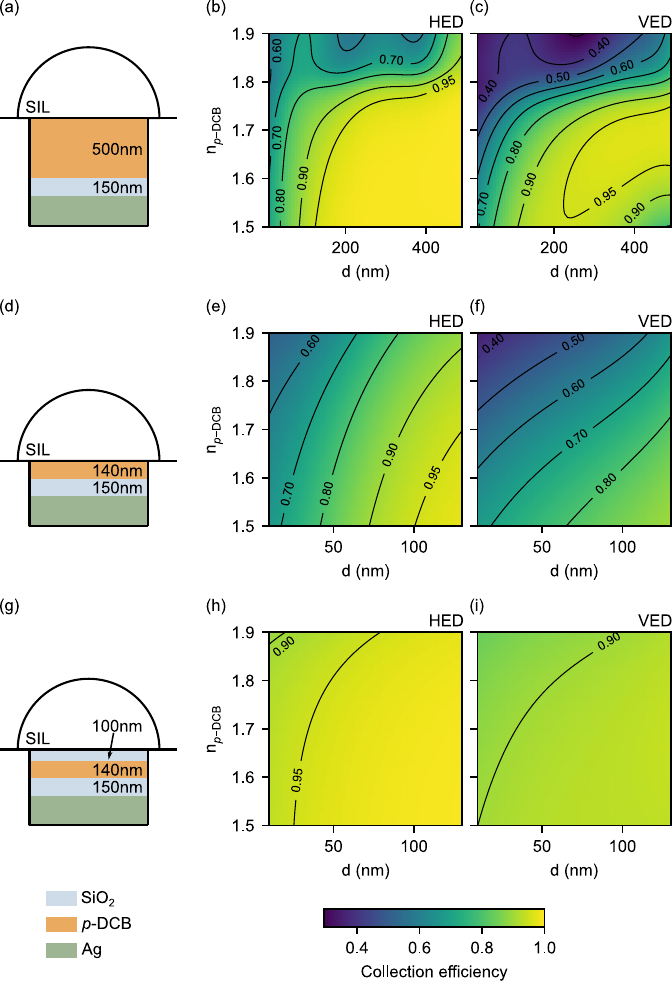}
    \caption{\textbf{Antenna designs and simulated photon collection efficiency.}
    Each row corresponds to one antenna design; the left column shows the layer structure, and the center and right columns show the simulated photon collection efficiency for a HED and a VED, respectively, as a function of the refractive index of the organic layer and the emitter distance $d$ from the SIL (or, in the third design, from the additional spacer) interface. The collection efficiency is evaluated for the collection lens with $\text{NA}=0.77$.
    (a)--(c) First design, featuring a 500-nm-thick \textit{p}-DCB layer;
    the large effective mode index limits the collection efficiency (b,c).
    (d)--(f) Second design, with the organic layer reduced to $140\,\mathrm{nm}$; the collection efficiency exceeds $95\,\%$  and $80\,\%$, respectively (e,f) but drops as the emitter approaches the high-index SIL interface (small $d$).
    (g)--(i) Third design, with an additional $100\,\mathrm{nm}$ SIL spacer between the organic layer and the SIL; the spacer suppresses this position dependence, maintaining a collection efficiency above $90\,\%$ across a broad range of emitter positions and refractive indices (h,i).}
    \label{collection_efficiency_colormaps}
\end{figure}

Next, using the analytical multilayer model \cite{chen2007efficient,chen201199}, we design and simulate the performance of three complete antenna structures, shown in Fig.~\ref{collection_efficiency_colormaps}. Each row corresponds to one design, with the layer structure shown in the left column and the corresponding simulated collection efficiency for a HED (middle column) and a VED (right column) shown as a function of the organic layer's refractive index and the emitter's distance $d$ from the SIL (or, in the third design, from the additional spacer) interface. 

In the first design (Fig.~\ref{collection_efficiency_colormaps}(a)), we use a $500\,\mathrm{nm}$ layer of \textit{p}-DCB. From the collection-lens side, the stack thus consists of the cubic zirconia SIL, a 500-nm-thick organic quasi-waveguide layer, a 150-nm-thick SiO$_2$ layer separating the organic layer from the silver mirror, which we refer to as the mirror spacer. For an emitter at the center of the organic layer and a refractive index of $1.9$, the collection efficiency drops below $60\,\%$ for both dipole orientations (Figs.~\ref{collection_efficiency_colormaps}(b),(c)), since the correspondingly large effective mode index of the lowest-order modes shifts the peak emission beyond the acceptance angle of the collection lens. High collection efficiency is thus confined to a broad region in the lower-right part of the maps, corresponding to low refractive index (bottom) and large emitter distance from the SIL (right). Throughout this region, the collection efficiency exceeds $90\,\%$ for both dipole orientations; in particular, values above $95\,\%$ are reached for any refractive index below $1.7$ as long as HED is located more than $200\,\mathrm{nm}$ from the SIL interface.

In the second design (Fig.~\ref{collection_efficiency_colormaps}(d)), we reduce the organic layer to $140\,\mathrm{nm}$, lowering the effective mode index safely below the requirement established above and ensuring a single-mode operation. This, however, brings a second condition for near-unity collection efficiency into play: the emitter must be located at a distance of about $\lambda/(2n)$ from the high-index interface, where $\lambda = 744\,\mathrm{nm}$ is the vacuum wavelength of the molecular emission and $n$ is the bulk refractive index of the film. Otherwise, the evanescent components of the dipole radiation can efficiently couple to the continuum of modes in the high-index layer, resulting in emission angles that exceed the critical angle and are thus lost to collection. Comparing Figs.~\ref{collection_efficiency_colormaps}(e),(f) shows that this constraint is more pronounced for the VED than for the HED. For the HED, the collection efficiency remains above $70\,\%$ as long as the molecule is more than $70\,\mathrm{nm}$ from the SIL interface, even at a refractive index as high as $1.9$, while for the VED it reaches about $50\,\%$ under the same conditions. In both cases, the efficiency increases as the refractive index of the organic layer decreases.

In the third design (Fig.~\ref{collection_efficiency_colormaps}(g)), we address the loss due to evanescent coupling by inserting a second, thin $100\,\mathrm{nm}$ SiO$_2$ spacer -- the SIL spacer -- between the organic layer and the SIL, in addition to the mirror spacer already present between the organic layer and the silver mirror. This SIL spacer increases the effective distance between the emitter and the high-index interface, preventing molecules from coupling into the unwanted large-angle modes regardless of their exact position within the crystal. The calculations show that a thickness of $100\,\mathrm{nm}$ is already sufficient: the collection efficiency exceeds $90\,\%$ over a broad range of emitter positions and refractive indices, for both dipole orientations (Figs.~\ref{collection_efficiency_colormaps}(h),(i)), thereby strongly reducing the sensitivity to both the emitter position and the refractive index of the organic layer. We note, however, that the collection efficiency of the VED is in general about $5\,\%$ lower than that of the HED, since its near-field couples more strongly to the metal and thus leads to stronger Ohmic absorption in the silver layer \cite{chen2007efficient,chen201199}.

\begin{figure}[!htbp]
   \centering
   \includegraphics{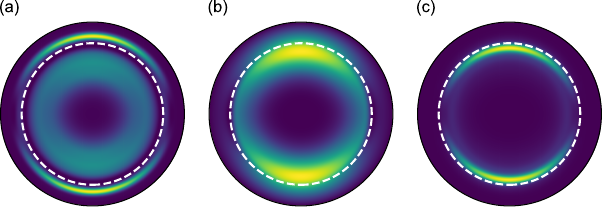}
    \caption{\textbf{Simulated far-field radiation patterns for the three antenna designs.}
Far-field emission of a HED placed at the center of the organic layer, computed by FDTD for a fixed, isotropic refractive index of $n=1.9$.
(a) First design ($500\,\mathrm{nm}$ organic layer).
(b) Second design ($140\,\mathrm{nm}$ organic layer, no SIL spacer).
(c) Third design ($140\,\mathrm{nm}$ organic layer with an additional $100\,\mathrm{nm}$ SIL spacer between the organic layer and the SIL).
The white dashed circle marks the collection angle corresponding to the aspheric lens with $\text{NA}=0.77$. Emission outside this circle is not captured in the experiment.}
    \label{fig_farfield}
\end{figure}

Above, we introduced the three designs and quantified the collection efficiency as a function of the refractive index and the emitter position. While those maps provide the quantitative picture, the underlying mechanism is most easily appreciated in the far-field emission patterns obtained from FDTD calculations. Figs.~\ref{fig_farfield}(a)--(c) display the emission pattern of a horizontally emitting dipole placed at the center of the organic layer for all three designs, shown here as a representative example. The white dashed circle in each panel marks the collection angle corresponding to $\text{NA}=0.77$. For the 500-nm-thick layer (Fig.~\ref{fig_farfield}(a)), the large effective mode index of the lowest order mode directs the emission to angles beyond the collection cone. The thin layer instead produces a single lowest-order mode for HED, but near-field coupling to the high-index substrate still redirects a substantial fraction of the emission outside the usable NA (Fig.~\ref{fig_farfield}(b)). The third design remedies both shortcomings, confining the emission almost entirely within the collection cone (Fig.~\ref{fig_farfield}(c)).

We note that the far-field patterns in Fig.~\ref{fig_farfield} were computed for a fixed, isotropic refractive index of $1.9$, corresponding to the largest principal index of \textit{p}-DCB. In reality, \textit{p}-DCB is biaxial, with principal refractive indices of approximately $1.9$, $1.64$, and $1.45$~\cite{manghi1967optical}. Since two of the three principal axes exhibit a substantially lower index, this fixed value represents a conservative worst case, realized only when both the corresponding crystal axis and the molecular transition dipole are aligned with the relevant direction. Since the way in which DBT incorporates into the \textit{p}-DCB lattice is not known, the resulting effective index cannot be predicted a priori; however, whenever the transition dipole projects predominantly onto the lower-index axes, the effective index approaches $1.45$--$1.64$, corresponding to the favorable regime spanned by the lower part of the efficiency maps in Fig.~\ref{collection_efficiency_colormaps}, where high collection efficiencies are obtained even for the thicker organic layer.

Among the three designs, the third offers the most robust performance, maintaining a collection efficiency above $90\,\%$ quite independent of the emitter position, the orientation of the molecule within the crystal, and the exact refractive index of the organic layer. Its only drawback is that the additional SIL spacer requires one further deposition step. Since this step integrates naturally into the fabrication workflow described below, we nonetheless regard the full antenna as a fairly straightforward extension. Here, we focus on the simpler second design as a first demonstration.

\section{Sample fabrication}

\begin{figure}[!htbp]
    \centering
    \includegraphics{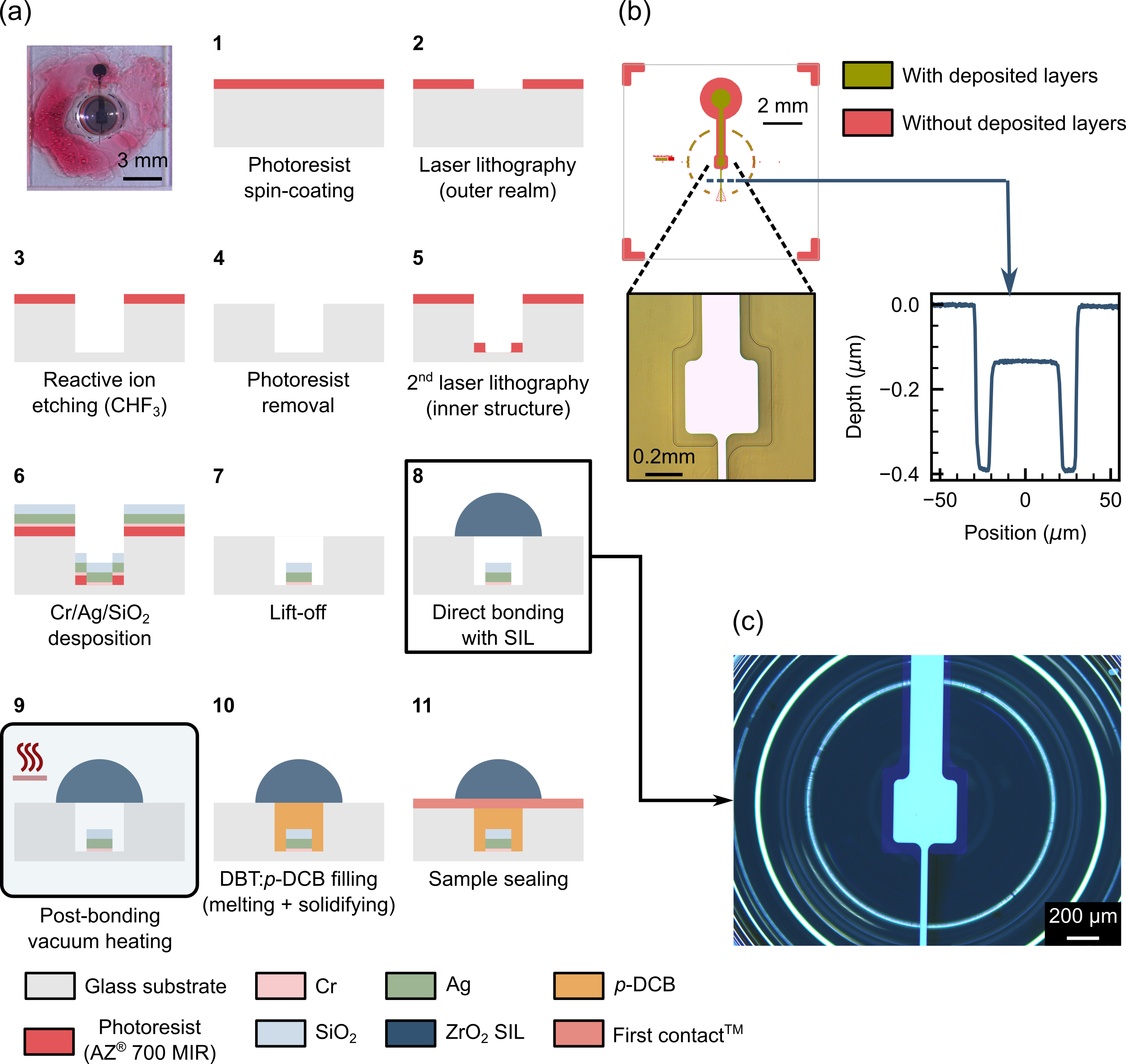}
    \caption{\textbf{Fabrication of the metallo-dielectric antenna incorporating organic materials.}
(a) Photograph of the finished, sealed antenna sample (top-left panel), together with a schematic overview of the fabrication procedure; the individual process steps are numbered 1--11 and referenced in the main text. 
(b) Schematic of the patterned substrate and a microscope image of the central structure, where the deposited Ag/SiO$_2$ stack appears as the strongly reflective central region. The channel cross section shown on the lower right was obtained by profilometry along the exit channel, marked by the blue dashed line in the upper left panel.
(c) Microscope image of the antenna, viewed through the transparent substrate after bonding and prior to crystal filling; the absence of interference fringes outside the channel region confirms uniform bonding.}
    \label{fig_fabrication_overview}
\end{figure}

Fabricating a complete antenna structure is nontrivial, since the device combines organic and inorganic, as well as amorphous and crystalline materials within a single structure, each imposing its own processing constraints. In the following, we present a concrete fabrication workflow that accounts for the specific requirements of each material and yields the complete antenna structure, including the optional SIL spacer, a thin \ce{SiO2} layer between the SIL and the organic crystal, which can be included or omitted depending on the desired design. The fabrication process, summarized in Fig.~\ref{fig_fabrication_overview}(a), proceeds in three main phases. First, the antenna substrate is fabricated using a top-down approach based on two lithography cycles, producing a precisely defined trench together with the deposited layers in the central area; the latter include the \ce{SiO2} mirror spacer between the silver mirror and the organic layer, which is present in all designs. Second, the SIL, optionally coated with the SIL spacer on its flat side, is bonded directly to the etched substrate, yielding a solid one-piece assembly whose semi-enclosed channel later serves as the site for crystal growth. Third, the organic material is melted, introduced into the channel, solidified in a controlled manner, and finally sealed to prevent evaporation during vacuum pumping.

The nanostructured substrate, composed of \ce{Ag} and \ce{SiO2} layers, provides the reflective and low-index components of the antenna. As illustrated in steps 1--7 of Fig.~\ref{fig_fabrication_overview}(a), it is fabricated using a dual-lithography protocol combining etching and thin-film deposition. The resulting etched pattern, highlighted in Fig.~\ref{fig_fabrication_overview}(b), features a circular crystal-loading inlet, a rectangular region of interest beneath the center of the SIL, and a narrow exit channel. Together, these elements guide the molten crystal into the target region during the subsequent growth step. Along the vertical axis, the channel cross-section has a central depth of $140\,\text{nm}$ and a side-trench depth of $400\,\text{nm}$, as confirmed by profilometry (Alpha-Step D-500, KLA; Fig.~\ref{fig_fabrication_overview}(b)). This height difference results from the sequential deposition of the \ce{Cr} adhesion layer, the \ce{Ag} mirror, and the $150\,\text{nm}$ \ce{SiO2} mirror spacer within the central domain, which together define the antenna's waveguide geometry: the $\sim\!260\,\text{nm}$ total thickness of this deposited stack reduces the $400\,\text{nm}$ side-trench depth to the $140\,\text{nm}$ central channel. Deposition is deliberately excluded from the surrounding area to avoid sharp ridges during lift-off, since residual material accumulated at the channel boundaries would otherwise degrade both surface flatness and bonding quality.

To implement this design, the first lithography cycle (steps 1--4) defines the channel with a uniform depth of $400\,\text{nm}$. A glass substrate is spin-coated with photoresist (AZ\textsuperscript{\textregistered} 700 MIR, Merck GmbH) and patterned by laser lithography (DWL 66+, Heidelberg Instruments), covering both the red- and green-shaded regions in Fig.~\ref{fig_fabrication_overview}(b); the exposed areas are then etched with a \ce{CHF3} plasma (PlasmaPro100 Cobra, Oxford Instruments) to the target depth. After photoresist removal, a second lithography cycle (steps 5--7) exposes only the region corresponding to the inner structural features, onto which the \ce{Cr}, \ce{Ag}, and \ce{SiO2} layers are deposited by sputtering (ATC Orion-5, AJA International). Here, the \ce{Cr} layer serves as an adhesion promoter for the subsequent \ce{Ag} mirror and \ce{SiO2} mirror spacer. A final lift-off step removes all excess material, leaving only the deposited stack within the central channel, which appears as the strongly reflective central region in the microscope image of Fig.~\ref{fig_fabrication_overview}(b).

For designs incorporating the SIL spacer, this layer would be deposited onto the planar surface of the SIL, prior to bonding. Atomic layer deposition (ALD) is well suited for this purpose, as it provides the required nanometer thickness control and excellent uniformity across the SIL surface. In the present work, however, we realize the simpler design without the SIL spacer, so that the bare SIL surface is bonded directly to the substrate, as described next.

%For designs incorporating the SIL spacer, this layer can be deposited at this stage onto the planar surface of the SIL, prior to bonding, using atomic layer deposition (ALD; TFS200, Beneq) with a target thickness of $100\,\text{nm}$; the substrate temperature is raised to $150\,^\circ\text{C}$ during deposition to improve adhesion. The thickness and refractive index of the deposited layer are characterized by ellipsometry (Auto SE, Horiba) at five evenly distributed positions across the surface, with values and uncertainties obtained from curve fitting of the ellipsometry data. The resulting average thickness is $103.0\,\text{nm}$, with a sample-to-sample deviation of at most $3.2\,\text{nm}$ and a within-sample variation below $0.5\,\text{nm}$, confirming excellent thickness uniformity both within and across SILs. This step is omitted entirely for designs without the SIL spacer, in which case the bare SIL surface is bonded directly to the substrate, as described next.

Following substrate preparation, the second phase (steps 8--9) attaches the SIL to the substrate. We depart here from the mechanical clamping used for SIL mounting in previous designs~\cite{rattenbacher2019coherent,zirkelbach2022high}, which allowed the molten crystal to infiltrate the SIL--substrate interface and form a crystalline layer several micrometers thick. As discussed in Sec.~\ref{sec_simulation}, such a thick crystal layer raises the effective index of the quasi-waveguide modes and lowers the collection efficiency. The problem is further compounded by the formation of multiple crystal domains through the channel depth, which introduce additional scattering at the domain boundaries.

Direct bonding via intermolecular interactions offers a reliable alternative and is widely used for this purpose in the semiconductor industry \cite{plossl1999wafer}. Robust adhesion requires both surfaces to be free of contaminants: the SIL is cleaned in a sequential solvent bath of Extran\textsuperscript{\textregistered}, water, and isopropanol, and both components are subsequently treated with a gentle \ce{O2} plasma (Atto, Diener Electronic) to render their surfaces hydrophilic. The resulting hydroxyl groups mediate hydrogen bonding upon room-temperature contact. To realize the bonding, the SIL is positioned slightly above the center of the substrate using a vacuum pick-up tool (VP10C, Thorlabs), and once the vacuum is released, the lens settles onto the substrate. Afterwards, gentle pressure is applied to its top surface to enhance interfacial adhesion. This yields an assembly that is stable enough for handling yet still readily separable by external forces or solvents. Optical microscopy of the bonded sample (Fig.~\ref{fig_fabrication_overview}(c)) confirms that uniform bonding is already established at this stage, as evidenced by the absence of interference fringes outside the channel region. To further strengthen the bond, we anneal the sample at elevated temperature, a technique well established for bonded silicon wafers.  The sample is heated to $150\,^\circ\text{C}$ for 5--10 hours under vacuum (VACUTherm, Thermo Scientific), yielding a structure with excellent mechanical and chemical robustness.

In the final stage (steps 10--11), the organic crystal is grown inside the defined channel to form a well-ordered crystalline layer. We first prepare the guest-host system by doping DBT molecules into \ce{\textit{p}-DCB} at a concentration of approximately $0.02\,\text{ppm}$; combined with our antenna geometry, this concentration restricts resonant excitation to at most one molecule within a $20\,\mathrm{\mu m}$-wide illumination area.

\begin{figure}[!htbp]
    \centering
    \includegraphics{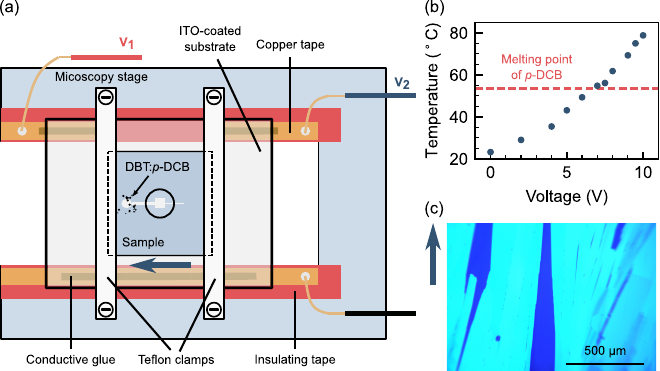}
    \caption{\textbf{Crystal growth inside the channel.} (a) Sketch of the crystal-growth platform. Spatial control of the temperature is achieved by tuning two independent voltages applied to opposite sides of the ITO-coated substrate. (b) Temperature measured at the center of the ITO plate as a function of the applied voltage, calibrated with a thermal camera; the melting point of \textit{p}-DCB is indicated by the red dashed line. (c) Cross-polarized microscope image of the \textit{p}-DCB crystal inside the channel. Distinct crystal domains, corresponding to different crystallographic orientations, appear as regions of different color. The blue arrow in (b) and (c) represents the direction of crystal growth.}
    \label{fig_crystal_growth}
\end{figure}

Capillary-driven filling of the molten crystal into the shallow channel requires surface activation. The bonded structure is therefore treated with a 5-minute \ce{O2} plasma to promote wetting. The sample is then transferred to a home-built crystal-growth setup, sketched in Fig.~\ref{fig_crystal_growth}(a), for controlled solidification. The organic sample is melted via Joule heating of an indium tin oxide (ITO)-coated glass substrate (Sigma-Aldrich). The slide is securely attached to three copper strips using conductive epoxy, in which one strip runs the full length of the slide and connects to the negative terminal of a DC power supply (E3620A, Keysight), while two additional strips on the opposite side, electrically isolated from each other, connect to independent positive terminals. This arrangement provides spatial control of the temperature along the channel axis: uniform heating is obtained for $V_1=V_2$, with the absolute temperature set by the applied voltage, while $V_1\neq V_2$ instead generates a controlled thermal gradient along the channel. The entire assembly is secured on a microscopy stage, with insulating tape placed beneath the copper contacts to prevent electrical shorting.

The temperature--voltage relationship was calibrated using a thermal camera (Fig.~\ref{fig_crystal_growth}(b)). Guided by this calibration, the crystal-growth protocol proceeds as follows. The ITO-coated slide is first preheated for about 5 minutes and then held slightly above the melting point of \ce{\textit{p}-DCB} ($54\,^\circ\text{C}$). The bonded sample is then secured to the heating stage with Teflon clamps, chosen for their low thermal conductivity. Consequently, a small amount of the DBT:\ce{\textit{p}-DCB} mixture is deposited around the channel entrance using a spatula. Owing to the preheating, the material rapidly liquefies and wicks into the channel by capillary action within seconds. Both heating voltages are then lowered below the melting threshold, after which the voltage on one side is further reduced to establish a temperature difference of approximately $6\,^\circ\text{C}$ between the two ends of the channel. This thermal gradient is found to yield the smoothest, most directional crystal growth. The effectiveness of this protocol is evident in the microscopy image of Fig.~\ref{fig_crystal_growth}(c), where the resulting domains are elongated along the growth direction and extend over hundreds of micrometers. To our knowledge, the resulting 140-nm-thick organic crystal is the thinnest uniformly melt-crystallized film reported to date for single-molecule experiments.

Typically used host crystals for molecular emitters exhibit high vapor pressure and are prone to sublimation during vacuum pumping, making reliable sealing essential. Vacuum grease (Apiezon N) is sometimes used for this purpose owing to its vacuum compatibility~\cite{zirkelbach2022high}, but its tendency to contaminate surfaces and resist removal poses significant fabrication challenges. We instead use First Contact\textsuperscript{TM} as the sealing medium -- visible as the red covering layer in the top-left panel of Fig.~\ref{fig_fabrication_overview}(a). Originally formulated for optical cleaning, this polymer rapidly cures into a conformal seal that can later be cleanly peeled from optical surfaces with negligible residue. This sealing step concludes the fabrication workflow, yielding a sample ready for transfer into a cryogenic environment.

\section{Back-focal-plane imaging of a single molecule inside the antenna at cryogenic temperature}

Having fabricated the antenna in its simpler form, without the SIL spacer, we now characterize it at cryogenic temperature. The sample was cooled to $1.4\,\mathrm{K}$, and individual DBT molecules were characterized by BFP imaging. Because the back focal plane of the objective maps emission angle onto radial position, a BFP image encodes the angular radiation pattern and is therefore sensitive to both the dipole orientation $\theta$ and the emitter depth $d$ within the channel -- the same two parameters that govern the collection efficiency in our design.

\begin{figure}[!htbp]
    \centering
    \includegraphics{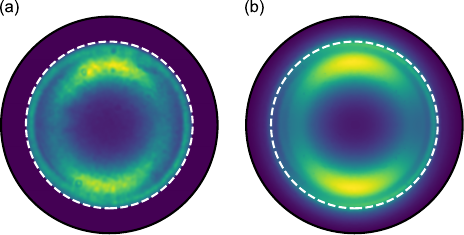}
    \caption{\textbf{Extraction of dipole properties from BFP imaging.} (a) Measured BFP image of a single molecule inside the antenna. The white dashed circle marks the lens's NA as used in the experiment. Values outside are set to zero, as this region falls outside the lens's acceptance angle. (b) FDTD simulation of a dipole located inside the antenna at a distance of $90\,\mathrm{nm}$ from the SIL interface, oriented at an angle of $\theta = 60^\circ$ relative to the horizontal plane.}
    \label{fig_bfp_fitting}
\end{figure}

Acquiring complete BFP images at cryogenic temperatures is challenging because cryo-compatible objectives have limited NAs, restricting the collection of high-angle emission that carries critical far-field information. Fig.~\ref{fig_bfp_fitting}(a) shows a BFP image of a single DBT molecule obtained under these conditions, chosen as a representative example. The measured pattern exhibits features consistent with those expected from an emitter embedded inside the antenna. However, since the emitter's exact position, its dipole orientation, and the orientation of the birefringent organic crystal within the channel are all a priori unknown, a quantitative fit to the data is not straightforward: each combination of position, dipole orientation, and crystal orientation requires a distinct FDTD simulation, which makes an exhaustive search computationally prohibitive. We therefore compare the measurement against a representative set of simulations spanning different emitter positions and dipole orientations. To simplify the problem of the unknown crystal orientation, we fix the effective refractive index for the two dipole components.

As a first step, we use the position of the inner lobes -- attributable to the horizontal dipole component -- to determine an effective refractive index experienced by that component, thereby constraining one degree of freedom. For the vertical component, however, the corresponding feature -- its outer-ring maximum -- lies at large angles and may be truncated by the NA accessible in our setup. As we cannot tell whether, and how much of, this maximum is cut off, the refractive index experienced by the vertical component cannot be determined with the same confidence but needs to be estimated, which makes it difficult to identify the matching simulation unambiguously. 

Starting at a distance of $10\,\mathrm{nm}$ from the SIL spacer, we simulate horizontal- and vertical-dipole radiation patterns in $20\,\mathrm{nm}$ depth increments, using refractive indices of $1.75$ for the horizontal and $1.65$ for the vertical dipole component. Superimposing both contributions with varying weights, we then construct radiation patterns for a range of dipole orientations, from purely in-plane to fully out-of-plane, in $10^\circ$ steps. The simulated pattern that best matches the experimentally measured BFP image serves as our reference.

The result is shown in Fig.~\ref{fig_bfp_fitting}(b): we find good agreement for a dipole orientation of $\theta = 60^\circ$ and an emitter depth of $d = 90\,\mathrm{nm}$ from the SIL interface, confirming that the fabricated antenna reproduces the expected radiation pattern. For this configuration, the simulation predicts a collection efficiency of $79\,\%$. We emphasize that this collection efficiency is inferred from the best-matching simulated configuration rather than measured directly. It nonetheless represents a first step toward a full experimental characterization of the collection efficiency, which we leave to future work.

\section{Conclusion}
In summary, we have established design and fabrication principles for cryogenic metallo-dielectric antennas optimized for single organic molecules. Our design for near-unity photon collection is based on a layered dielectric stack terminated by a metallic reflector, which redirects emission toward the collection optics, combined with a thin organic layer that keeps the effective mode index of the quasi-waveguide modes -- and hence the peak emission angle -- within the acceptance cone of a lens with $\text{NA}=0.77$. Incorporating an additional SIL spacer that suppresses evanescent coupling between the emitter and the high-index substrate is particularly appealing, as it promises collection efficiencies exceeding $90\,\%$ for arbitrary dipole orientation at this NA, while remaining broadband and tolerant to the emitter's position within the organic layer.

Such an antenna could, in the future, also enable the determination of an emitter's quantum efficiency at cryogenic temperature via a simple saturation measurement -- a quantity that has so far been difficult to determine reliably at low temperature, since the collection efficiency entering such a measurement depends sensitively on the emitter's dipole orientation~\cite{musavinezhad2023quantum}, which is itself difficult to establish without access to the back-focal plane.

On the fabrication side, we demonstrated a robust route to this novel architecture for metallo-dielectric antennas: direct bonding of a solid immersion lens to a nanostructured channel, followed by growth of a thin, uniformly melt-crystallized organic film. Back-focal-plane imaging of single \ce{DBT} molecules shows radiation patterns consistent with those expected from our design, indicating that the fabricated structure -- realized here in its simpler form without the SIL spacer -- operates as intended; incorporating the spacer layer into the fabrication workflow, following the design and process steps outlined above, remains an extension for future work.

This approach is not specific to \ce{DBT}:\ce{\textit{p}-DCB} and should transfer to other guest-host systems, providing a scalable, cryo-compatible platform for bright and coherent single-photon sources. Future work will focus on precisely quantifying the collection efficiency of an antenna with a SIL spacer, and on employing it in quantum-optical and quantum-information applications.

\begin{backmatter}
\bmsection{Funding}
The authors acknowledge financial support from the Max Planck Society, the Deutsche Forschungsgemeinschaft (DFG, German Research Foundation) -- ID 429529648 -- TRR 306 QuCoLiMa (“Quantum Cooperativity of Light and Matter”) and the Free State of Bavaria via the Munich Quantum Valley light house project ”QuMeCo”.

\bmsection{Acknowledgment}
We would like to thank Vahid Sandoghdar for his continuous support and valuable discussions. We would also like to thank the TDSU1 team at Max Planck Institute for the Science of Light for their assistance with the fabrication of the antenna.

%\bmsection{Acknowledgment}
%Additional information crediting individuals who contributed to the work being reported, clarifying who received funding from a particular source, or other information that does not fit the criteria for the funding block may also be included; for example, ``K. Flockhart thanks the National Science Foundation for help identifying collaborators for this work.'' 

\bmsection{Disclosures}
The authors declare no conflicts of interest.

% \medskip

% \noindent ABC: 123 Corporation (I,E,P), DEF: 456 Corporation (R,S). GHI: 789 Corporation (C).

% \medskip

% \noindent If there are no disclosures, then list ``The authors declare no conflicts of interest.''

\bmsection{Data Availability Statement}
Data underlying the results presented in this paper are not publicly available at this time but may be obtained from the authors upon reasonable request.

\end{backmatter}

\bibliography{sample_3}

%%%%%%%%%% If preparing manually:
% \begin{thebibliography}{1}
% \newcommand{\enquote}[1]{``#1''}

% \bibitem{Zhang:14}
% Y.~Zhang, S.~Qiao, L.~Sun, Q.~W. Shi, W.~Huang, L.~Li, and Z.~Yang,
%   \enquote{Photoinduced active terahertz metamaterials with nanostructured
%   vanadium dioxide film deposited by sol-gel method,}
%   {\protect\JournalTitle{Optics Express}} \textbf{22}, 11070--11078 (2014).

% \bibitem{Optica}
% {Optica}, \enquote{{Optica Publishing Group},}
%   \url{http://www.opg.optica.org}.

% \bibitem{FORSTER2007}
% P.~Forster, V.~Ramaswamy, P.~Artaxo, T.~Bernsten, R.~Betts, D.~Fahey,
%   J.~Haywood, J.~Lean, D.~Lowe, G.~Myhre, J.~Nganga, R.~Prinn, G.~Raga,
%   M.~Schulz, and R.~V. Dorland, \enquote{Changes in atmospheric consituents and
%   in radiative forcing,} in \enquote{Climate Change 2007: The Physical Science
%   Basis. Contribution of Working Group 1 to the Fourth Assesment Report of
%   Intergovernmental Panel on Climate Change,}  S.~Solomon, D.~Qin, M.~Manning,
%   Z.~Chen, M.~Marquis, K.~B. Averyt, M.~Tignor, and H.~L. Miler, eds.
%   (Cambridge University Press, 2007).

% \end{thebibliography}

\end{document}